# Who shapes crisis communication on Twitter? An analysis of influential German-language accounts during the COVID-19 pandemic

Gautam Kishore Shahi, Sünje Clausen and Stefan Stieglitz
University of Duisburg-Essen, Germany
(gautam.shahi,sünje.clausen and stefan.stieglitz)@uni-due.de

## Abstract

*Twitter is becoming an increasingly important platform for disseminating information during crisis situations, such as the COVID-19 pandemic. Effective crisis communication on Twitter can shape the public perception of the crisis, influence adherence to preventative measures, and thus affect public health. Influential accounts are particularly important as they reach large audiences quickly. This study identifies influential German-language accounts from almost 3 million German tweets collected between January and May 2020 by constructing a retweet network and calculating PageRank centrality values. We capture the volatility of crisis communication by structuring the analysis into seven stages based on key events during the pandemic and profile influential accounts into roles. Our analysis shows that news and journalist accounts were influential throughout all phases, while government accounts were particularly important shortly before and after the lockdown was instantiated. We discuss implications for crisis communication during health crises and for analyzing long-term crisis data.*

## 1. Introduction

Over the last decade, social media has evolved into an important tool for procuring and disseminating information in different domains [1]. The low transaction costs to exchange information, in addition to the significantly higher usage of mobile devices over the last years, further increased the rates in which ad-hoc information is shared [2]. As individuals use social media platforms as an information source during crises [3, 4], they present an opportunity for the government and health organizations to quickly disseminate crisis-relevant information among the public and shape public perception of the crises. Thereby, the communication of crisis managers and stakeholders can influence individual perceptions of the crisis positively or negatively [5], and the trust of the population depends on the measures taken by these actors [6]. Crisis communication can also influence the perceived risk or threat posed by a crisis and thereby influence the adherence to preventative measures and ultimately the safety of society [7]. Misleading information on preventative measures shared on social media can influence behavior and cause fatal injuries if the information is inaccurate [8]. Thus, understanding crisis communication on social media can support managing crises more effectively [9] by enabling health professionals, researchers, and policymakers to protect public health better, improve policy decisions, and provide quality health information [10].

In the ongoing health crisis caused by the coronavirus disease (COVID-19) outbreak, Twitter became a key communication, and public health dissemination tool which provides the opportunity to influence the health of large audiences positively [11]. In Germany, the popularity of Twitter is continuing to rise as 22% of German internet users were active Twitter users in 2020, compared to only 7% back in 2015 [12]. For example, the German police uses Twitter for informing the public, with some accounts having up to half a million followers (e.g., @polizeiberlin; September 2021). Previous research shows that Twitter was used for collective sense-making during a terrorist attack in Berlin, Germany [13]. Thus, Twitter is an increasingly important platform in Germany to disseminate and gather information, especially during crises. Thereby, some Twitter users have a particularly big influence on the network and reach a large audience. The content which these influential accounts share is distributed in micro-networks on the platform and therefore receives a lot of attention from many people [14]. In the case of the COVID-19 pandemic, the national, state, and local governments of different countries had to implement preventative measures to inhibit the spread of the virus, such as washing hands, wearing face masks, and social distancing. Here, influential accounts could raise awareness of health issues and initiate changes at the societal level [15, 16] which could improve the crisis response during pandemics. Identifying influential accounts and improving the understanding of crisis communication could be decisive for preventing or attenuating future COVID-19 infection waves [17].

However, crises can be complex, often involve different stakeholders, and evolve over time. As the crisis changes, so do crisis communication. One option to consider the dynamic course of crisis communication in the analysis is to separate the communication into several timeframes [18]. Previous research categorized crises into multiple stages [19, 20] which capture the different behaviors and emotions that individuals might exhibit and experience depending on the current development of the crisis. For example, before (or at the beginning of) a crisis, there might be higher uncertainty, perceived risk, and associated information-seeking behavior than in later crisis stages. Furthermore, as the COVID-19 pandemic evolved differently in different countries and local authorities responded differently to the threat, an investigation of the connection between the development of the pandemic and public communication must consider geographical factors. Therefore, like Park, Park, and Chong [21], who analyzed the COVID-19 conversation in Korea, this research focuses on German-language communication. To summarize, the social media platform Twitter is an important platform for crisis communication and could support (or hinder) effective crisis response. Thereby, influential accounts have a larger impact on the content which gets distributed in a network and can thus shape crisis communication. Lastly, the stage of the crisis and geographical location are relevant boundary conditions potentially influencing crisis communication. Against this background, we pose the following research question:

*RQ: What types of German-language accounts were influential on Twitter during the different phases of the COVID-19 pandemic?*

To answer the research question, we analyzed 3,026,543 tweets collected during the COVID-19 pandemic by structuring them into phases according to key events, and the stages of a crisis [19]. We constructed a retweet network for each phase and calculated PageRank values to determine the most influential accounts. These accounts were then manually reviewed and profiled into roles. Our research contributes to practice by illustrating which types of accounts were among the most influential in the public debate during the COVID-19 pandemic, which could support the evaluation of past and the management of future crises [18, 9]. Furthermore, we contribute to research by presenting a structured approach for analyzing volatile crisis communication based on key events and in relation to the stages of a crisis. The remainder of the paper is structured as follows: We review relevant literature before presenting our methodological approach to identify influential accounts in the different phases of the COVID-19 pandemic. Lastly, results, limitations, and possible extensions of this work are presented and discussed.

## 2. Literature Background

In this section, we discuss relevant previous work on crisis communication, social media analytics, influential accounts in crisis communication and provide a brief overview of the development of the COVID-19 pandemic in Germany.

### 2.1. Crisis Communication

In severe crises, such as natural disasters, pandemics, or terrorist attacks, the public often experiences uncertainty and negative emotions [22]. The crisis communication of the media, government, organizations, and among citizens can shape the perception of the crisis, and behavioral responses to the threat [6]. For example, the perceived risk or threat of the crisis can influence the willingness and capability of individuals to adhere to protective behavioral measures [7]. Thereby, crisis communication changes in accordance with the development of the crisis. To account for the dynamic course of crisis communication, the analysis can be separated into several timeframes [18]. For example, Chew and Eysenbach used four weeks period to analyze the Twitter communication regarding misinformation during the H1N1 pandemic of 2009 [23]. However, the selection of these timeframes tends to be arbitrary concerning the development of the crisis itself. The development of a crisis was formalized by Fink [19], who structured crises into four stages according to certain characteristics and key events. In the prodromal stage (1) first signs of a possible crisis can be observed. A key event then initiates the crisis breakout stage (2). In the chronic stage (3), the effects of the crisis are visible over a longer period and despite taking measures. Finally, in the resolution stage (4), stakeholders and the public no longer consider the situation to be threatening. In this research, we combine the described approaches in the analysis of crisis communication by selecting the time frames following the stages of the crisis, as defined by key events.

### 2.2. Social media analytics for crisis communication

Social media is used to communicate with family, friends, and acquaintances to share content, exchange opinions, and obtain information. In times of uncertainty, individuals use Social Media platforms to communicate with each other, gather information, and make sense of the situation [24]. Primarily Twitter and Facebook evolved into important mediums for crisis communication both as an information source and a communication channel [4, 25]. Thereby, social media has become an important tool to disseminate (health) information among the public during crises but also to draw inferences on the public perception and response to disease outbreaks by using social media analytics [26]. A recent study showed that the organizational account is also involved in crisis communication, and around 53% of them spread harmful content [27].

### 2.3. Influencing crisis communication

As information spreads more effectively on Social Media if it is shared by accounts that are influential within the network [15, 16], identifying such accounts on Social Media marks an important step towards understanding crisis communication on Social Media and the role of government and health organizations in it. There are different definitions for influential accounts or "influencers" on social media [28, 29]. Previous research [30] has described influencers as "individuals accounts who disproportionately impact the spread of information or some related behavior of interest" (p. 2). Here, we use the term influential accounts to refer to accounts that disproportionately impacted the spread of information in the COVID-19 debate on Twitter.

Influence can be measured and quantified, for example, by considering the number of followers of an account (i.e., the popularity of an account) as it determines how many people will see the post. However, if the post's viewers are passive recipients, the information will not disseminate further through the network [31]. Therefore, an important measure is the level of engagement of Social Media users with the shared information by an account and their number of followers, as the interactions of users strongly impact further dissemination of the information within the network [31]. On Twitter, retweets, replies, and mentions are forms of engagement [32, 33] and especially retweets as an act of sharing content generated by other users increase the range and visibility of the information [34]. Thus, measures for influence increasingly analyze the account under consideration and the associated retweet network. A well-known measure that considers network information to quantify the influence of an account is eigenvector centrality [35]. Researchers argue that it is a suitable value in measuring node influence since this approach considers how influential the accounts retweeted the message's original message [36]. A variation of this measure is the PageRank algorithm which was specifically developed for directed graphs (such as retweet networks) and integrating the flow of influence to compute the authority of a node in a network [37]. Originally developed by Google and used to rank websites in Google search [38], it is applied for various purposes such as measuring influence in Social Networks [16]. For instance, it has been applied on graphs of retweets and replies for the analysis of influencers on Twitter [39]. Furthermore, compared to eigenvector centrality, the degree centrality of the neighbors is added up and divided by the number of successors. This limits the value of a node by its importance within the neighborhood [40]. Using a dampener for each hop in the researched nodes' neighborhood, the effect of direct and indirect influence decreases. Therefore, in this research, PageRank is chosen to measure the influence of the accounts.

### 2.4. COVID-19 in Germany

COVID-19 is an infectious respiratory disease that first appeared in Wuhan, China, in December 2019 [41]. It quickly spread in China and other countries around the world and caused an ongoing global health crisis. In Germany, the first cases were confirmed from January 28, 2020 [42]. To slow the spread of the disease, the German government recommended the cancellation of major events on March 9, closed schools from March 16, and instantiated a nationwide lockdown with physical contact restrictions ("social distancing") between March 23 and April 19. Since then, restrictions were partially loosened and are re-instantiated locally in response to rising infection numbers. To date (June 15, 2021), 3,716,170 Germans were tested positive, and 89,937 people died from COVID-19. Besides the detrimental health implications, the pandemic also increased unemployment by 14 % from March to April alone [43] and is still causing severe economic cuts, for example, for restaurants, the performing arts, and local businesses. The Kiel Institute for the world economy estimated at the end of April that economic performance in Germany dropped by about 20% during the lockdown [44]. Regardless of these negative impacts, it is likely many of the restrictions will persist until a critical proportion of the population is vaccinated. Thereby, effective crisis communication could improve adherence to preventative measures and public opinion of the crisis management.

## 3. Research Method

In this study, we have used the social media framework [2] to guide the data collection and analysis of Twitter data. The social media analytics process is visualized in Figure 1. In the following section, we provide a detailed description of the data collection and method used to identify and profile influential accounts.

| Phase | Timeframe (in 2021) | Short Description | Stage of crisis[19] | Top Accounts* |
|---|---|---|---|---|
| 1 | 01/28–02/26 | First infections started in some region of Germany | Breakout | DasErste, maischberger |
| 2 | 02/27–03/08 | of infection in other federal states | Breakout | jenshealthde, Fischblog |
| 3 | 03/09–03/22 | Two weeks before lockdown announced | Chronic | ZDFheute, heutejournal |
| 4 | 03/23–04/05 | First half of lockdown in Germany | Chronic | landnrw, tagesschau |
| 5 | 04/06–04/19 | Second half of lockdown in Germany | Chronic | tagesthemen,tagesschau |
| 6 | 04/20–05/03 | The government started first relaxations | Chronic | c_drosten, ZDFheute |
| 7 | 05/10–05/22 | Further relaxations and the situation normalized | Resolution | zuckerguss1998, JHillje |

Table 1. Division into multiple crisis phases with their respective names, the categorization, according to Fink [19], and the corresponding number of tweets within the dataset (*top two German accounts from each phase)

## 3.1. Data Collection

We have collected the data between January 28 and May 22, 2020, by applying a self-developed Java crawler using the Twitter4J library and saving it in a MySQL database. Since the terms and expressions used to refer to the COVID-19 pandemic changed over time, three separate crawling processes were conducted. First, worldwide German-speaking tweets were collected from January 28 to May 22, 2020, by tracking the following hashtags: #Coronavirus, #nCoV2019, #WuhanCoronaVirus, #WuhanVirus, #CoronaVirusOutbreak, #Ncov2020, and #coronaviruschina. Since the coronavirus got an official name by the WHO in February, new hashtags and terms appeared in the public discourse, which should be involved in the tracking process in order to retrieve a reliable data set. Thus, a second data crawling from February 27 to May 22, 2020, was conducted by tracking tweets using the hashtags: #covid19, #covid-19, #covid_19, #sarscov2, #covid, #cov and #corona. Furthermore, globally trending hashtags which were also used in German-language tweets, were crawled from March 8 to May 22, 2020: "corona Germany", #COVID19de, #coronapocalypse, #staythefuckhome, #flattenthecurve, and #stopthespread. Finally, we filtered the tweets in the German language, and all three crawled data sets were merged, and duplicates were removed. Overall, we have collected 310,366,886 multilingual tweets and obtained 3,026,543 German tweets for analysis for this study.

## 3.2. Data Processing

For the phase-wise analysis, the data set needed to be divided into different subsets. Considering the overall situation of COVID-19 in Germany, we divided the time span into different phases [45] following the key events provided by the Federal Ministry of Health [46]. Each phase was classified according to Fink's four-stage model [19]. The prodromal stage occurred prior to the start of the tracking, as signs of the pandemic could already be observed in other countries prior to the first infections in Germany. The phases and their classification are shown in table 1. The first period covers the recording of the first infection in Germany until it spreads to other federal states, indicating the breakout of a crisis. Therefore, it can be compared to the second stage of Fink's model [19]. Both subsequent periods reflect the phase before the lockdown in Germany and thus are essential phases in which communication increases and changes due to the rising level of perceived dread [47]. At the time of the lockdown, we define two phases as Jones et al. [48] imply that there are several peaks within a lockdown during short-term crisis communication, which we also want to investigate for long-lasting crises.

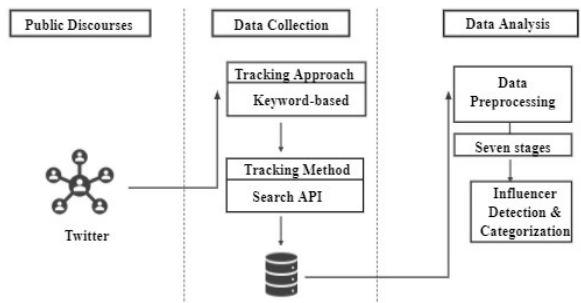

Figure 1. The framework used for analysis of public discourse on Twitter.

Furthermore, previous research [48] showed the occurrence of a relaxation phase after a lockdown, which was also investigated in this study. In Germany, this relaxation of the lockdown took place in two steps. In the first step, the lockdown was partly revoked, so that small shops and restaurants were allowed to open under certain conditions like mask obligation and distance regulation. In a second step, it was left to the federal states to carry out further relaxation, which is why there are differences in timing. During this second relaxation, larger shops and service providers such as hairdressers were also allowed to work with restrictions. This phase can be compared to

| Phase | Number of Tweets | Unique Accounts | Retweet Count* | Nodes | Edges | Degree* | Network Diameter | Clustering Coefficient* |
|---|---|---|---|---|---|---|---|---|
| 1 | 92,306 | 33,206 | 47.71 | 25,942 | 50,128 | 1.932 | 18 | 0.013 |
| 2 | 86,326 | 37,941 | 269.57 | 31,282 | 51,269 | 1.693 | 16 | 0.011 |
| 3 | 471,293 | 143,377 | 341.37 | 124,130 | 16,627 | 2.551 | 35 | 0.017 |
| 4 | 810,539 | 185,003 | 144.09 | 147,282 | 462,369 | 3.139 | 26 | 0.021 |
| 5 | 604,138 | 148,030 | 139.87 | 116,202 | 327,625 | 2.819 | 29 | 0.022 |
| 6 | 547,048 | 140,679 | 94.02 | 108,755 | 310,093 | 2.851 | 32 | 0.020 |
| 7 | 336,518 | 105,999 | 158.89 | 84,702 | 208,153 | 2.457 | 32 | 0.018 |

Table 2. Descriptive information for the data in the seven phases (*represents the mean value).

Fink's fourth stage, the resolution stage, as the perceived threat and uncertainty reduced and some normality was re-established.

Based on the division into subsets described above, further preprocessing and analysis was carried out. Each preprocessing and analysis step was conducted on each phase. Descriptive information about the analyzed data from each phase is provided in table 2. Influential accounts were identified based on a retweet network [49]. Using R in RStudio [50], the retweet network was constructed, resulting in a directed graph with nodes representing users who retweeted at least once or who retweeted at least once. The retweets are represented by edges from the retweeting user to the author of the original tweet. Nodes that do not have an edge at all are considered noisy and therefore extracted from the network.

### 3.3. Analysis of influential accounts

To identify influential accounts, we evaluated the user's PageRank score in a retweet network. In doing so, the most influential 100 accounts were identified in each phase using Gephi and their implemented version of PageRank with a probability of .85, epsilon of .001, and considering edge weights. Consequently, the accounts were profiled into roles in order to gain further insights into who shapes the Twitter debate. According to [51], who identified and categorized important Twitter users during the swine flu debate, influential accounts were classified into multiple categories. Similar to Stieglitz et al. [4], who identified roles within the sensemaking process during a crisis, we partially adopted information sources such as government, a public person (journalist), a public person (politician), a public person (celebrity), other public person and private person as roles. Additionally, the categories news, health professional, and science-related accounts were adapted from McNeill, Harris, and Briggs[52] who categorized users into pandemic relevant categories. Moreover, we added the category organization (non-health-related) and organization (health-related). This resulted in 11 distinct categories: science-related account (e.g., virologist), organization (non-health-related), organization (health-related), health professional (e.g., medical doctor), private person, public person (celebrity; e.g., musician, actor, entertainer), public person (politician), public person (journalist), other public person, government, news. If a user fitted into more than one role, the content of the tweets was examined, and the category was chosen according to the predominating content of the tweets. The annotation process for roles included three annotators, and the labeling was conducted by examining the Twitter accounts and considering further information from the internet. First, two annotators independently labeled the accounts, and the results were compared. Disagreements were discussed until a consensus was reached. In the first round, the interrater reliability of .643 was computed using Cohen's kappa [53]. Subsequently, a coding book was specified in more detail in order to achieve a more consistent annotation. A third annotator used the adapted coding scheme to annotate the accounts, and these results were compared with the consensus reached by the first two annotators. The resulting Cohen's kappa of .777 signaled an acceptable to good interrater reliability. To unify the categorization of the first and second rounds, deviations in classification were discussed until consensus was reached.

### 4. Results

We provide the descriptive information of the included accounts and tweets, along with the social network statistics (number of nodes, edges, average degree, network diameter, and clustering coefficient) to provide more insights into the analyzed data per phase in table 2. With the spread of infections in Germany, communication on the topic increased. Within each phase, we identified the top 100 influential accounts by evaluating the user's PageRank scores. As mentioned before, these accounts were subsequently divided into different categories to give further insight into influential German-language accounts shaping the communication on Twitter during the COVID-19 pandemic. Out of 700 accounts, eight accounts were removed from the record because they had been deleted or were suspended by Twitter due to policy violations at the time of labeling.

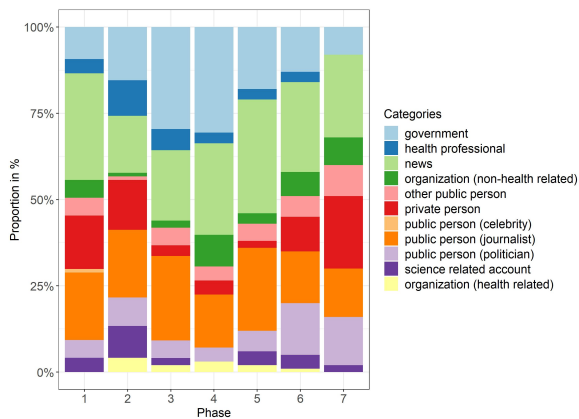

Figure 2. Graphical representation of the distribution of influential accounts in the seven phases.

Figure 2 provides an overview of the identified influential accounts in the seven stages of the pandemic. The results show that most of the identified accounts belong to the categories news, public person (journalist), and government. Of these three categories, the impact of the government accounts varied as there were less influential governmental accounts in the first and last phases. In contrast, however, the influential private persons are more often present in these two phases and less often in the intermediate ones. Here, private persons don't include explicitly to the celebrities. In Figure 3, the development of the influence of these accounts is illustrated. Thereby, news accounts and those of journalists were combined.

Another aspect that is noticeable in the distribution is that health-related accounts are generally not very often present, have a peak in the second phase, and decrease afterward. Only one influential celebrity was identified in the first phase, and no other celebrities were identified in other phases. No remarkable observations were found for the influential accounts of the other categories, which remained relatively constant throughout the phases.

## 5. Discussion

The identification and categorization of the influential accounts delivered a useful insight into the dynamics of the accounts which affect the public debate most during the different stages of the crisis. Overall, the number of tweets was highest shortly before and during the lockdown and started to decrease when the restrictions were loosened (Table 2). One reason could be the people were more concerned and had a higher demand for information during these early phases of the COVID-19 pandemic [27]. Similar to the research of Szomszor, Kostkova, and St Louis [54], who analyzed the preferences of trusted or untrusted sources on Twitter during the swine flu pandemic, we found that posts from news related accounts which are seen as trusted sources (categories: news and public person (journalist)) were generally popular during the predefined phases. However, contrary to the findings during the swine flu pandemic, health-related users were only influential during the breakout stage and had little influence in the other phases. This might be explained by the fact that when the virus started spreading in different federal states, COVID-19 was still relatively unknown, and professional opinions of health-related users were sought to meet the demand for information on respiratory disease. With time, the influence of these health-related users decreased as the novelty of news about the disease itself decreased, and more official, periodic information about the development of the disease was established (e.g., the daily press briefings by the Robert-Koch Institute).

As previously mentioned, the seemingly alternating impact of governmental accounts and private persons on the network was particularly interesting. The incline of influence of governmental accounts shortly before and in the first half of lockdown might be due to the associated uncertainty within the population. As the federal government is considered as one of the most trusted sources[55] individuals might have turned more frequently to government-related accounts in response to the uncertainty surrounding the nationwide lockdown. Also, in times of uncertainty, people seek to compensate for their lack of information by consuming news, especially from traditional media, as they are perceived to disseminate verified information from expert sources [56]. Since both traditional media and experts are represented on social media with their own respective accounts, it is likely for information seekers to consume the contents published by these sources, which explains the high number of news accounts and journalists among the influential accounts.

The finding that the influence of health-related (and later also governmental) accounts decreases within the ongoing chronic stage of the pandemic might also be an indicator of fatigue due to the dominance of COVID-19-related information and the adherence to preventive measures. The negative effects of, for example, social distancing on individual well-being could result in the population neglecting social distancing practices before the government relaxes or reverses the adopted measures [57].

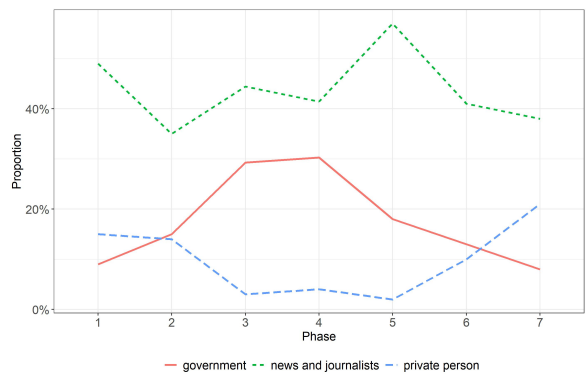

Figure 3. Distribution of the groups news & journalists, government, and private persons. Phase 3 covers the two weeks before the lockdown, while phases 4 and 5 cover the time of the lockdown itself.

To continue containing the disease, awareness and adherence to preventive measure is key. Therefore, we suggest exploring new and more proactive communication strategies for health-related information and a stronger consideration of individual well-being in crisis communication. First, as news and journalists consistently rank high in influence throughout the pandemic, health organizations should consider reaching out proactively to media accounts and journalists to disseminate their content. In light of the finding that private accounts regain influence in the chronic stage of the pandemic and in accordance with Bavel et al. [22], health organizations could also increase their efforts to identify and reach out to influential private accounts who are, for example, religious or community leaders to different audiences to share public health messages. This is also supported by Mirbabaie et al. [58] who emphasized the importance of information-rich actors and their influence on communication networks to facilitate sense-making during crises. Second, we suggest considering the emotional and mental well-being of the population more carefully in crisis communication by acknowledging and discussing, for example, feelings of loneliness. Potentially, the population could be advised on "COVID-19-compatible" strategies. For example, Block et al. [59] show first promising approaches with the concept of contact groups, which should limit possible infection chains to fewer individuals [59]. Lastly, the low proportion of celebrities among the influential accounts was unexpected. In contrast to a short-term crisis, where celebrities are among the most influential due to a high number of retweets [18], they appear to have little or no relevance within the COVID-19 pandemic in German-language Twitter communication. This might be because celebrities were either not creating original content or were not considered relevant sources for health-related information. Approaching them might present an opportunity for more effective dissemination of public health messages. For example, in the UK, social media influencers were paid to promote test and trace services. Thereby, the types of influencers and messages should be evaluated carefully, as, for example, Singh et al. [60] showed that involving social media influencers in corporate crisis communication can harm reputation and trustworthiness if the intent is perceived to be manipulative.

By taking Fink's model [19] of crisis stages as a basis and extending it into multiple phases based on key events during the pandemic, we were able to derive more nuanced insights into the development of influential actors throughout the pandemic. Importantly, we found indicators that the influence of governmental accounts increases at the beginning of the chronic stage and drops after a rather short time frame. These peaks would not have been found by the standard procedure of a breakdown into the crisis stages by Fink (1986) [19] and probably also not by dividing the data set into arbitrary timeframes of equal size (e.g.,[61]). For future research of long-term crises such as pandemics, we, therefore, propose to structure the data and analysis according to significant key events within the crisis stages.

## 6. Conclusion and Future work

In the present study, we showed that the highest proportion of influential German-speaking Twitter users during the COVID-19 pandemic were news and journalist accounts and the second most impactful were governmental accounts. While the news-related accounts were strongly present in all seven predefined phases, the governmental accounts were especially influential shortly before and at the beginning of the lockdown but not as much at the beginning of the pandemic (phase 1) and the later chronic stages (phase 6). Contrary to the governmental users, the private accounts were especially influential in the first and last phase. We discussed possible strategies to account for these changes in influence to improve crisis communication and adherence to preventive measures during long-term crises such as pandemics.

There are several limitations of this work. First, the number and type of relevant tweets changed throughout the data acquisition process, for example, due to new descriptors of the virus emerging. The tracking focused on universally used hashtags; however, including more local, German-language hashtags might have provided additional relevant data. Further, the Twitter API did not allow exhaustive tracking of such extensive communication, and the number of collected tweets and retweets differed between the phases (see table 2). Thus, while almost 3 million tweets were used in total, our analysis might not be representative of the entire Twitter communication on COVID-19 in the respective phases. And, of course, Twitter communication only represents a part of German society. Second, as the data collection was language-based and not geolocation-based, tweets from other German-language countries and international individuals tweeting in German are included in the data set as well. These tweets might not have referred to the course of the pandemic in Germany. For example, while Switzerland shows a similar time pattern, the governmental measures in Austria were taken at a slightly different time than in Germany and Switzerland. Recent literature shows that the selection of the period under investigation can significantly influence the results [4]. Therefore, data from Austria might have skewed the analysis. Lastly, the sampling and methodological approach for quantifying influence in a network influences obtainable results. Thus, the insights on the different phases are an approximation of the evolution of influential accounts. In future research, the proposed analysis approach can be applied in other countries to detect potential differences in influential actors during long-term crises. Especially in areas with different numbers of infections, the comparison of communication can yield interesting findings. The analysis could also be enriched by using topic modeling or sentiment analysis to provide insights into how users feel about the virus and governmental measures in different pandemic stages. Furthermore, to verify our results regarding the low number of influential celebrities, actor-based tracking approaches could be used to analyze celebrities' contents and communication behavior during pandemics. Lastly, one could analyze whether the proportion of misinformation changed in the different stages of the pandemic, how rumors developed and affected collective sense-making [25].